\documentclass[a4paper,11pt,dvitex]{article}
\usepackage{pos}

\def\simge{\mathrel{%
       \rlap{\raise 0.511ex \hbox{$>$}}{\lower 0.511ex \hbox{$\sim$}}}}
\def\simle{\mathrel{%
       \rlap{\raise 0.511ex \hbox{$<$}}{\lower 0.511ex \hbox{$\sim$}}}}

\title{Chemical potential dependence of the endpoint of first-order phase transition in heavy-quark region of finite-temperature lattice QCD}
\ShortTitle{Chemical potential dependence of the endpoint of first-order phase transition}

\author*[a]{Shinji Ejiri}
\author[b]{Kazuyuki Kanaya}
\author[c,d]{Masakiyo Kitazawa}

\affiliation[a]{Department of Physics, Niigata University, Niigata 950-2181, Japan}

\affiliation[b]{Tomonaga Center for the History of the Universe, University of Tsukuba, Tsukuba, Ibaraki 305-8571, Japan}

\affiliation[c]{Yukawa Institute for Theoretical Physics, Kyoto University, Kyoto 606-8502, Japan}

\affiliation[d]{J-PARC Branch, KEK Theory Center, Institute of Particle and Nuclear Studies, KEK, 203-1 Shirakata, Tokai, Ibaraki, 319-1106, Japan}

\emailAdd{ejiri@muse.sc.niigata-u.ac.jp}

\abstract{
We determine the location of the critical point where the first-order deconfining transition in the heavy-quark region turns into a crossover in finite-temperature and density lattice QCD with 2+1 flavors of Wilson quarks.
Combining a hopping parameter expansion (HPE) of the quark determinant with a reweighting method, we evaluate the chemical potential dependence of the critical point.
By systematically calculating the coefficients of the hopping parameter expansion up to a high order of HPE at finite chemical potential, we find that the higher order terms are strongly correlated with the Polyakov loop, which is the leading-order term, on each configuration.
Moreover, their complex phases themselves, which are important at finite density, are also found to be strongly correlated with the complex phase of the Polyakov loop.
Using this property, we develop a method for estimating the critical point incorporating high-order terms from calculations with only low-order terms.
We report that the first-order phase transition region in the heavy-quark region becomes narrower exponentially with increasing the chemical potential.
Since the hopping parameter of the critical point decreases exponentially as the density increases, the sign problem does not become serious even when the density increases, and critical points can be evaluated up to high densities.
}

\FullConference{The 40th International Symposium on Lattice Field Theory (Lattice 2023)\\
July 31st - August 4th, 2023\\
Fermi National Accelerator Laboratory\\}

\begin{document}
\maketitle

\section{Introduction}
\label{intro}

The phase structure of QCD depends not only on temperature and chemical potential but also on the quark mass.
In 2+1 flavor QCD, there are first-order phase transition regions around the heavy quark limit \cite{Saito:2011fs,Saito:2013vja,Ejiri:2019csa,Kiyohara:2021smr,Cuteri:2020yke} and also near the massless point of 3 flavors.
We have been investigating the critical line at the boundary where the first-order phase transition ends in the heavy quark region \cite{Saito:2011fs,Saito:2013vja,Ejiri:2019csa,Kiyohara:2021smr,Wakabayashi:2021eye,Kitazawa2023pos}.
When quark masses are heavy, approximation by the hopping parameter expansion (HPE) is useful.
This approximation drastically reduces computational demands, and allows us to use reweighting methods for quark masses, which is difficult in full QCD simulations.

In this report, we extend the study to the complex phase of the terms in the HPE at finite chemical potential.
We introduce the HPE in finite temperature/density lattice QCD in the next section.
We particularly discuss the effects of the chemical potential. 
In Sec.~\ref{sec:critical}, we determine the critical line when the complex phase is ignored, i.e., in the phase quenched QCD.
Then, in Sec.~\ref{sec:complex}, we estimate the effect of the complex phase on the critical line and calculate the shift of the critical line from the phase quenched case.
Section \ref{summary} is devoted to a summary.

\section{Hopping parameter expansion}
\label{sec:hopping}

We consider $N_{\rm f}$ flavor QCD with the standard plaquette gauge action and the Wilson quark action.
The expectation value of an operator ${\cal \hat{O}}$ is given by 
\begin{equation}
\langle {\cal \hat{O}} \rangle_{(\beta, \kappa)}
= \frac{1}{Z} \int {\cal D} U \, {\cal \hat{O}} \, \prod_{f=1}^{N_{\rm f}} \det M(\kappa_f) \, e^{6 \beta N_{\rm site} \hat{P}} , 
\hspace{5mm}
Z= \int {\cal D} U \, \prod_{f=1}^{N_{\rm f}} 
\det M(\kappa_f) \, e^{6 \beta N_{\rm site} \hat{P}} , 
\label{eq:expect}
\end{equation}
where $N_{\rm site}=N_s^3 \times N_t$ is the number of sites, $T=1/(N_t a)$ is the temperature, and $a$ is the lattice spacing.
The plaquette value is $\hat{P}$.
The Wilson quark kernel $M_{xy}$ is defined by  
\begin{eqnarray}
M_{xy} (\kappa) &=& \delta_{xy} 
- \kappa \left[ \sum_{j=1}^3 
\left\{ (1-\gamma_{j})\,U_{x, j}\,\delta_{y,x+\hat{j}} 
+ (1+\gamma_{j})\,U_{y, j}^{\dagger}\,\delta_{y,x-\hat{j}} \right\} 
\right. \nonumber \\ && \left.
+ (1-\gamma_4)\,U_{x, 4}\; e^{\mu a} \,\delta_{y,x+\hat{4}} 
+ (1+\gamma_4)\,U_{y, 4}^{\dagger}\; e^{-\mu a} \,\delta_{y,x-\hat{4}} ,
\right]
\label{eq:Mxydf}
\end{eqnarray}
with the chemical potential $\mu$ and the hopping parameter $\kappa$.
The hopping parameter is approximately proportional to the inverse quark mass.
In the heavy quark region, we evaluate the quark determinant by the HPE around $\kappa=0$:
\begin{eqnarray}
\ln \det M(\kappa) & = &
\ln \det M(0) + N_{\rm site} \sum_{n=1}^{\infty} D_{n} \kappa^{n} , 
\label{eq:tayexp} \\
D_n &=& \frac{1}{N_{\rm site} \ n!} \left[ \frac{\partial^n \ln \det M}{\partial \kappa^n} \right]_{\kappa=0}
\ = \ \frac{(-1)^{n+1}}{N_{\rm site} \ n} \; {\rm tr} 
\left[ \left( \frac{\partial M}{\partial \kappa} \right)^n \right] .
\label{eq:derkappa}
\end{eqnarray}
The first term is $\ln \det M(0) =0$, and
$(\partial M/\partial \kappa)_{xy}$ is the hopping term following $\kappa$ in the right hand side of Eq.~(\ref{eq:Mxydf}).
Nonzero contributions appear when the product of the hopping terms form closed loops in the space-time. 
Therefore, the nonzero contributions to $D_n$ are given by $n$-step Wilson loops and Polyakov loops. 
The latter are closed by the boundary condition, where we impose the anti-periodic boundary condition in the time direction for fermions.

Moreover, these $D_n$ are classified by the number of windings $m$ in the time direction:
\begin{eqnarray}
D_n &=& \hat{W}(n) + \sum_{m=1}^{\infty} \hat{L}_m^+ (N_t, n) e^{m \mu/T} + \sum_{m=1}^{\infty} \hat{L}_m^- (N_t, n) e^{-m \mu/T}
\nonumber \\
&=&  \hat{W}(n) + \sum_{m=1}^{\infty} 2 {\rm Re} \hat{L}_m^+ (N_t, n) \cosh \left( \frac{m \mu}{T} \right) 
+ i \sum_{m=1}^{\infty} 2 {\rm Im} \hat{L}_m^+ (N_t, n) \sinh \left( \frac{m \mu}{T} \right) .
\label{eq:loopex}
\end{eqnarray}
$\hat{W} (n)$ is the sum of Wilson-loop-type terms with winding number zero, and $\hat{L}_m^+ (N_t, n)$ is the sum of Polyakov-loop-type terms with winding number $m$ in the positive direction. 
$\hat{L}_m^- (N_t, n)$ is that with winding number $m$ in the negative direction, 
and $\hat{L}_m^- (N_t, n)=[\hat{L}_m^+ (N_t, n)]^*$.
$\hat{W}(n)$ is nonzero only if $n \geq 4$ and $n$ is even.
When $N_t$ is an even number, $\hat{L}_m^+ (N_t, n)$ is nonzero only if $n$ is an even number, and the lowest order term is $\hat{L}_m^+ (N_t, mN_t)$, i.e.,
$\hat{L}_m^+ (N_t, n)$ is nonzero when $n \geq m N_t$.

\begin{figure}[tb]
\centering
\vspace{-8mm}
\includegraphics[width=5.9cm,clip]{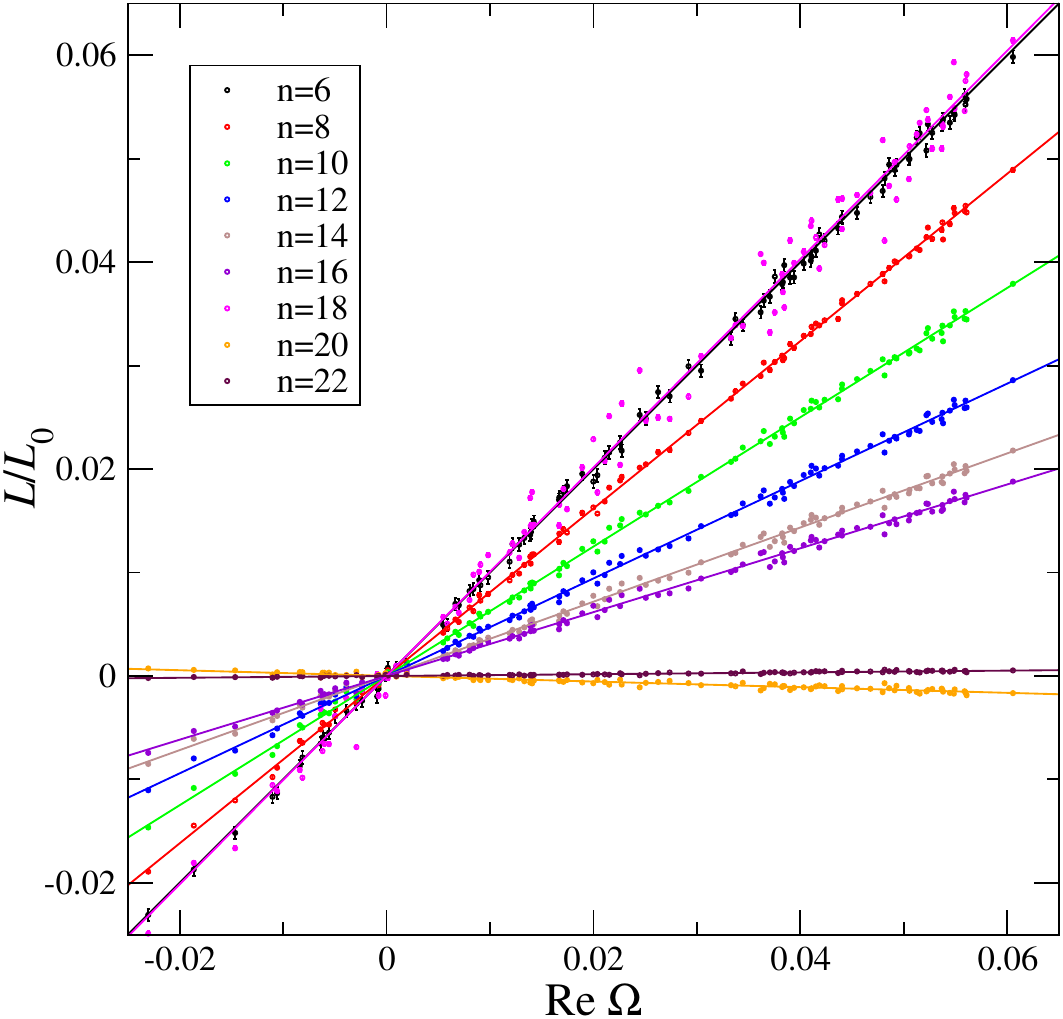}
\hspace{5mm}
\includegraphics[width=6.0cm,clip]{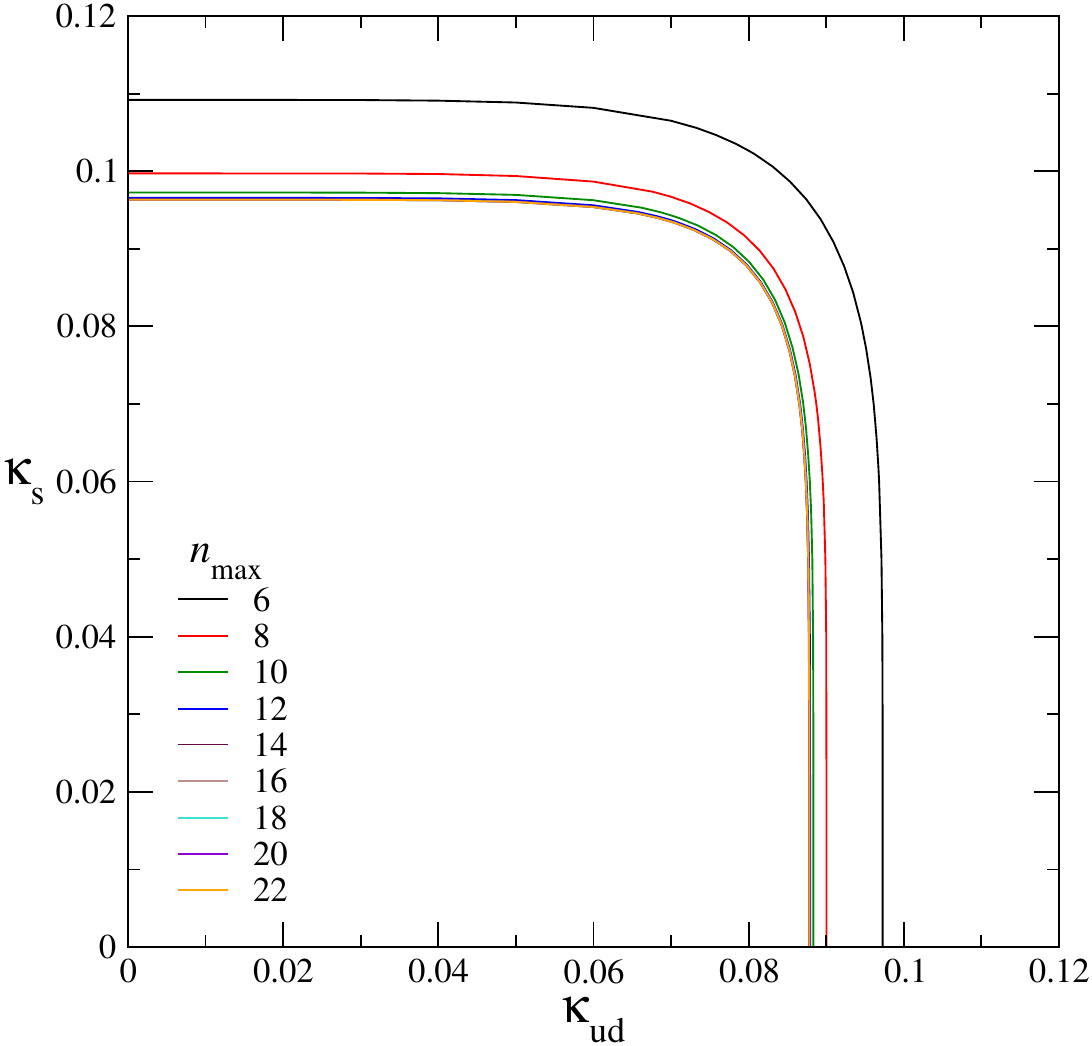}
\vspace{-1mm}
\caption{Left: Correlation between $\hat{L}(N_t, n)/L^0(N_t,n)$ and ${\rm Re} \hat{\Omega}$ for $N_t= 6$.
The straight lines are the fit functions.
Right: The critical $\kappa$ of 2+1 flavor QCD calculated with various $n_{\rm max}$ for $N_t= 6$.
}
\label{fig1}
\end{figure}

These expansion coefficients have been calculated on configurations generated near the phase transition point in Ref.~\cite{Wakabayashi:2021eye} and found to have the following properties: 
(1)~$\hat{W}(n)$ mainly effect to shift the gauge coupling $\beta$, and have almost no effects in the determination of the critical $\kappa$.
(2)~$\hat{L}_m (N_t, n)$ for $m \geq 2$ are much smaller than $\hat{L}_1 (N_t, n)$. 
(3)~$\hat{L} (N_t, n)$  is strongly correlated with ${\rm Re} \hat{\Omega}$ on each configuration, i.e.,
\begin{eqnarray}
\hat{L} (N_t, n) \approx L^0 (N_t, n) c_n {\rm Re} \hat{\Omega},
\label{eq:lnapp}
\end{eqnarray}
where $\hat{L} (N_t, n) = \sum_m [\hat{L}_m^+ (N_t, n)+ \hat{L}_m^- (N_t, n)] =  2\sum_m {\rm Re} \hat{L}_m^+ (N_t, n)$, $c_n$ is a proportionality constant, and $L^0(N_t, n)$ is $\hat{L}(N_t, n)$ when all link fields are set to $U_{x, \mu}=1$, which is given in Table 2 of Ref.~\cite{Wakabayashi:2021eye}.
Using these properties, we define an effective action:
\begin{eqnarray}
S_{\rm eff} = -6N_{\rm site} \beta^* \hat{P} - N_s^3 \lambda^* {\rm Re} \hat{\Omega} , 
\hspace{3mm} {\rm where} \hspace{3mm}
\lambda^* = N_t \sum_{f=1}^{N_{\rm f}} \sum_{n=N_t}^{n_{\rm max}} L^0(N_t, n) \cosh \left( \frac{\mu}{T} \right) \kappa_f^{n} c_n ,
\label{eq:seff}
\end{eqnarray}
with which the partition function is given by  
$Z= \int {\cal D} U \, e^{-S_{\rm eff}} \cos \theta$. 
Here, $\theta$ is the complex phase of the quark determinant, which is nonvanishing at finite $\mu$ and will be discussed in Sec.~\ref{sec:complex}.
$\hat{W}(n)$ effectively shift $\beta$ from the original $\beta$ to 
$\beta^* = \beta + (1/6) W^0(4) \sum_f \kappa_f^4+ \cdots$ \cite{Wakabayashi:2021eye}.
However, in the determination of the critical $\kappa$, $\beta$ is adjusted to the phase transition point for each $\kappa$, so we do not discuss the shift of $\beta$ in this report.
We plot ${\rm Re} \hat{\Omega}$ and $\hat{L}(N_t, n)/L^0(N_t, n)$ on each configuration in the left panel of Fig.~\ref{fig1}.
The configurations are generated at $\beta^*=5.8905$ and $\lambda^*=0.0012$ on a $36^3 \times 6$ lattice.
This figure shows that the data are linearly distributed and that Eq.~(\ref{eq:lnapp}) is well satisfied, at least up to the order $n=22$.
The slopes define $c_n$ and their values turned out to have only vary weak dependences on $\beta^*$ and $\lambda^*$.
When the approximation of Eq.~(\ref{eq:seff}) is valid with a finite $n_{\rm max}$, the computational cost can be drastically reduced, since the calculation of $\det M$ is not required in the simulation.

\section{Critical line for phase quenched QCD}
\label{sec:critical}

We performed Monte Carlo simulation using the action $S_{\rm eff}$
on $N_t =4$ and 6 lattices~\cite{Kiyohara:2021smr,Kitazawa2023pos}.
At the transition point $\beta_{\rm c}$ for each $\kappa$, we determined the critical point $\kappa_{\rm c}$ where the order of the phase transition changes from first-order to crossover.
We generated configurations with $S_{\rm eff}$ and calculated the operators $\hat{L}(N_t, N_t)$ and $\hat{L}(N_t, N_t+2)$.
We corrected the measurements to be accurate up to the next leading order (NLO) of HPE by taking the effects of $\hat{L}(N_t, n)$ up to $n=N_t+2$ with the reweighting method. Performing a scaling analysis of the order parameter ${\rm Re} \hat{\Omega}$, we determined the critical point $\kappa_{\rm c, NLO}$.
The results for $N_{\rm f}=2$ are $\kappa_{\rm c, NLO}=0.0602(4)$ at $N_t =4$ \cite{Kiyohara:2021smr} and $\kappa_{\rm c, NLO}=0.09003(19)$ at $N_t =6$~\cite{Kitazawa2023pos}.

When we adopt the effective theory of Eq.~(\ref{eq:seff}),  the critical point at $\mu=0$ and that at finite $\mu$ in the phase quenched QCD (i.e., ignoring the complex phase of $\det M$) are dependent only on the parameters $\beta^*$ and $\lambda^*$.  
The dependence on the number of flavors are inherent in these parameters through explicit relations by the HPE. 
Therefore, when the critical point for $N_{\rm f}=2$ is obtained, we can determine it for any $N_{\rm f}$  with no new simulations.
From $\kappa_{\rm c, NLO}$ of $N_{\rm f} =2$, the critical $\lambda^*$ is 
$\lambda^*_c = 2 N_t [ L^0(N_t, N_t) \kappa_{\rm c, NLO}^{N_t} 
+ L^0(N_t, N_t+2) \kappa_{\rm c, NLO}^{N_t+2} c_{N_t+2}]$.
For $N_{\rm f}=2+1$, denoting the hopping parameter for the up and down quarks as $\kappa_{\rm ud}$ and that for the strange quark as $\kappa_{\rm s}$, the critical line in the $(\kappa_{\rm ud}, \kappa_{\rm s})$ plane is obtained by finding $(\kappa_{\rm c, ud}, \kappa_{\rm c, s})$ that satisfies the following equation :
\begin{eqnarray}
2 \sum_{n=N_t}^{n_{\rm max}} L^0 (N_t, n) \, \cosh \left( \frac{\mu}{T} \right) c_n \kappa_{\rm c, ud}^n
+ \sum_{n=N_t}^{n_{\rm max}} L^0 (N_t, n) \, \cosh \left( \frac{\mu}{T} \right) c_n \kappa_{\rm c, s}^n 
= \frac{\lambda^*_c}{N_t} .
\label{eq:keff3f}
\end{eqnarray}

\begin{figure}[tb]
\centering
\vspace{-7mm}
\includegraphics[width=6.0cm,clip]{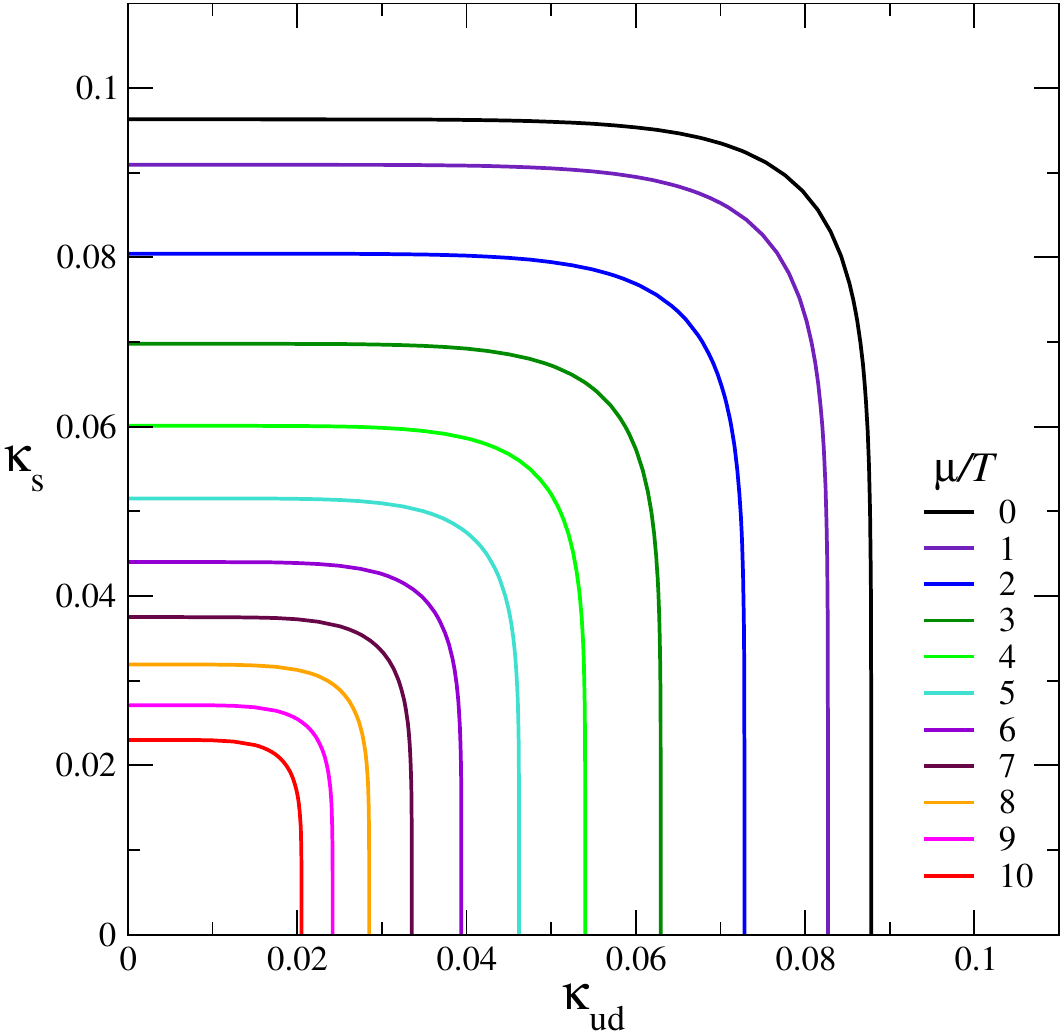}
\hspace{5mm}
\includegraphics[width=5.9cm,clip]{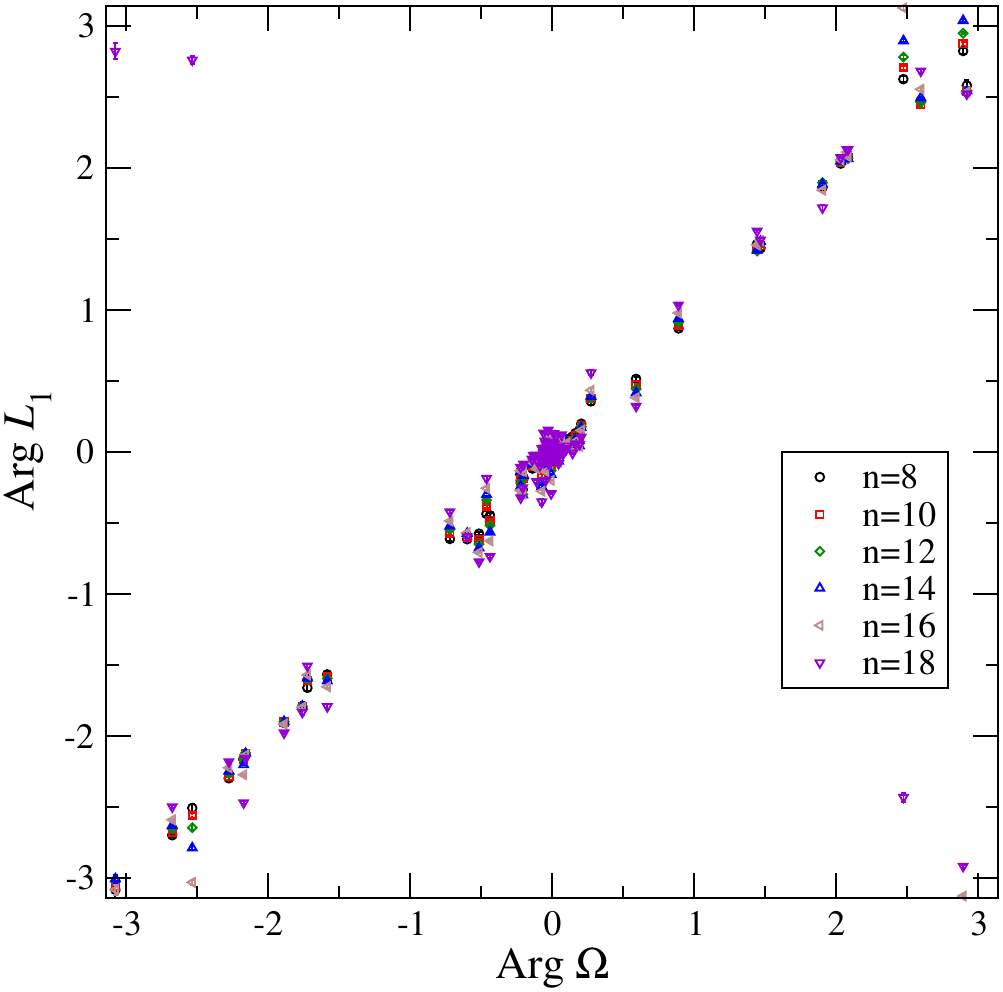}
\vspace{-1mm}
\caption{Left: The critical lines of 2+1 flavor  phase quenched finite density QCD  at each $\mu/T$ with $n_{\rm max}=22$.
Right: Correlation between ${\rm Arg} \hat{L}_1 (N_t, n)$ and ${\rm Arg} \hat{\Omega}$ for $N_t= 6$.
}
\label{fig2}
\end{figure}

The critical lines at $\mu=0$ for $N_t=6$ calculated with $n_{\rm max} = 6$--22 are shown in the right panel of Fig.~\ref{fig1}.
The region inside the critical line is the first-order phase transition region.
The critical line converges well when $n_{\rm max} \simge 10$. 
We also plot the critical lines of $N_{\rm f}=2+1$ phase quenched QCD  at each $\mu/T$ calculated with $n_{\rm max}=22$ in Fig.~\ref{fig2} (left).
This figure shows that the first-order phase transition region becomes exponentially narrower as $\mu$ increases.

\section{Complex phase of the quark determinant}
\label{sec:complex}

Next, we study the magnitude of the complex phase of $\det M$ on the critical point and calculate the shift of the critical point due to the complex phase.
The complex phase comes from the terms of $2 {\rm Im} \hat{L}_m^+ (N_t, n)$ in Eq.~(\ref{eq:loopex}).
We compute $\hat{L}_m^+ (N_t, n)$ on the configurations of $N_t=6$ used in Fig.~\ref{fig1} (left).
In order to calculate $\hat{L}_m^+ (N_t, n)$ and $\hat{L}_m^- (N_t, n)$ separately for $m = 1$ to 3, we impose eight types of boundary conditions,
$\psi(\vec{x}, N_t+1) = e^{2 \pi i k/8} \psi(\vec{x}, 1)$, 
in the temporal direction with $k=0$ to 7, and calculate the coefficients of the expansion using the noise method.
We plot the values of ${\rm Arg} \hat{\Omega} = \tan^{-1} ({\rm Im} \hat{\Omega}/{\rm Re} \hat{\Omega})$ and ${\rm Arg} \hat{L}_1^+ (N_t, n)$ for each configuration in Fig.~\ref{fig2} (right).
This figure indicates that the arguments of those complex numbers are approximately identical.
\begin{eqnarray}
{\rm Arg} \hat{L}_1^+ (N_t, n) \approx {\rm Arg} \hat{\Omega}, \hspace{3mm} {\rm thus,} \hspace{3mm}
2 {\rm Im} \hat{L}_1^+ (N_t, n) \approx L^0 (N_t, n) c_n {\rm Im} \hat{\Omega} .
\label{eq:argapp}
\end{eqnarray}
Note that Eq.~(\ref{eq:argapp}) is not satisfied for $n \geq 20$, since the sign of ${\rm Re} \hat{L}_1^+(6,n)$ changes to negative from $n=20$.
Regarding the complex phase of $\hat{L}_2^+ (N_t, n)$ and $\hat{L}_3^+ (N_t, n)$, since their values themselves are small, no clear signal could be obtained due to the error of the noise method.
We ignore the small contributions from $\hat{L}_2^+$ and $\hat{L}_3^+$ in the following.
From Eq.~(\ref{eq:argapp}), the complex phase of $\det M$ for degenerate $N_{\rm f}$ flavors is  given as follows, which is proportional to ${\rm Im} \hat{\Omega}$:
\begin{eqnarray}
\theta \approx N_{\rm f} N_{\rm site} 
\sum_{n=N_t}^{\infty} L^0 (N_t, n) \sinh \left( \frac{\mu}{T} \right) \kappa^n c_n {\rm Im} \hat{\Omega} .
\label{eq:tayexpapp}
\end{eqnarray}

In determining the critical point, we focus on the most important dynamical variable, the Polyakov loop $\hat{\Omega}$.
We define $w({\rm Re} \hat{\Omega})$ as the probability distribution function of ${\rm Re} \Omega$
when we generate configurations with the weight $|\det M|^{N_{\rm f}} e^{6 \beta N_{\rm site} P}$, 
and $\langle \cdots \rangle_{{\rm Re} \hat{\Omega} =x}$ as the expectation value averaged only when ${\rm Re} \hat{\Omega}$ is $x$.
Then, we rewrite Eq.~(\ref{eq:expect})  as 
\begin{eqnarray}
\langle {\cal O}[ {\rm Re} \hat{\Omega} ] \rangle =
\frac{1}{Z} \int {\cal O}[ {\rm Re} \hat{\Omega} ] \langle \cos \theta \rangle_{{\rm Re} \hat{\Omega} =x} w(x) dx ,
\label{eq:zpl}
\end{eqnarray}
where ${\cal O}[ {\rm Re} \hat{\Omega} ]$ is a physical observables in terms of ${\rm Re} \hat{\Omega}$, such as the susceptibility and the Binder cumulant of ${\rm Re} \hat{\Omega}$.
If the phase fluctuation is large, the sign problem occurs, 
i.e., $\langle \cos \theta \rangle_{{\rm Re} \hat{\Omega} =x}$ becomes zero within the error and the expectation value cannot be computed.
Even in such cases, it is possible to calculate the phase factor if the following cumulant expansion converges 
\cite{Ejiri:2009hq,Saito:2013vja}:
$\langle \cos \theta \rangle = \exp [ 1- \langle \theta^2 \rangle_c /2 
+ \langle \theta^4 \rangle_c /4! - \langle \theta^6 \rangle_c /6! + \dots ], $
where $\langle \theta^n \rangle_c$ are cumulants, e.g., 
$\langle \theta^2 \rangle_c = \langle \theta^2 \rangle$,
$\langle \theta^4 \rangle_c = \langle \theta^4 \rangle -3 \langle \theta^2 \rangle^2$,
$\langle \theta^6 \rangle_c = \langle \theta^6 \rangle 
-15 \langle \theta \rangle^4 \langle \theta^2 \rangle +30\langle \theta^2 \rangle^3$.
Since the exponential function is always positive, Eq.~(\ref{eq:zpl}) is computable.

\begin{figure}[tb]
\centering
\vspace{-8mm}
\includegraphics[width=6.6cm,clip]{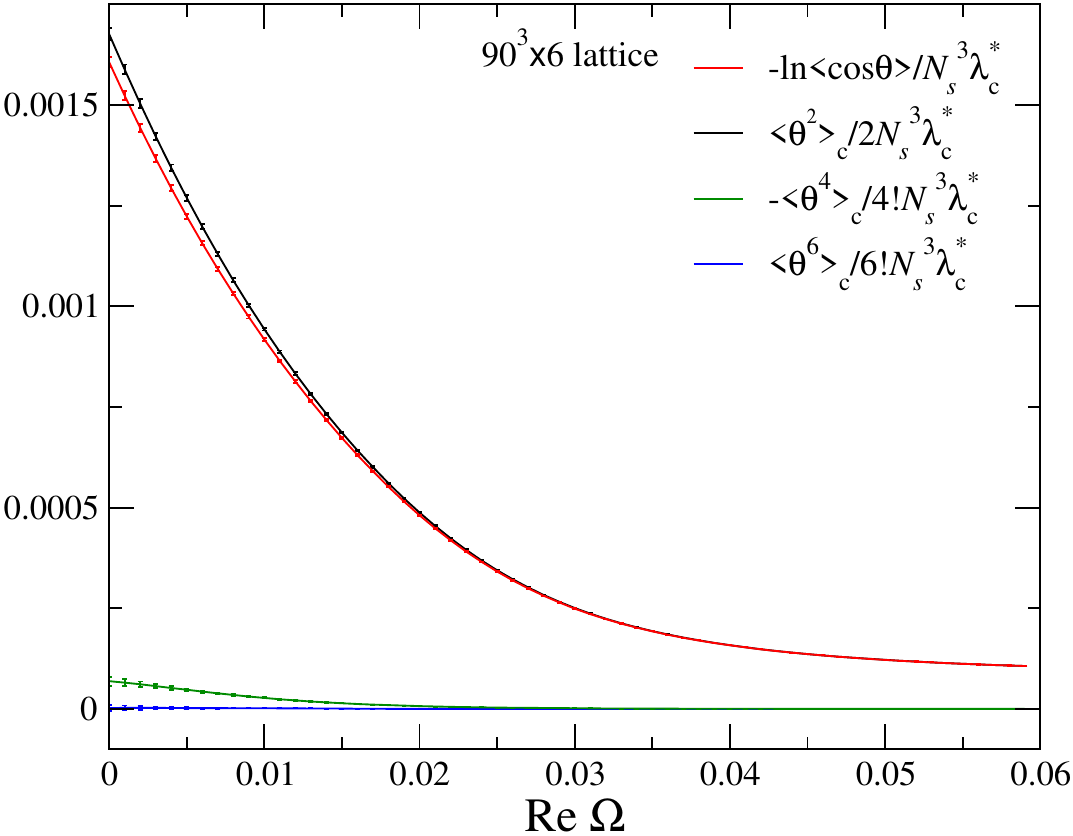}
\hspace{5mm}
\includegraphics[width=6.4cm,clip]{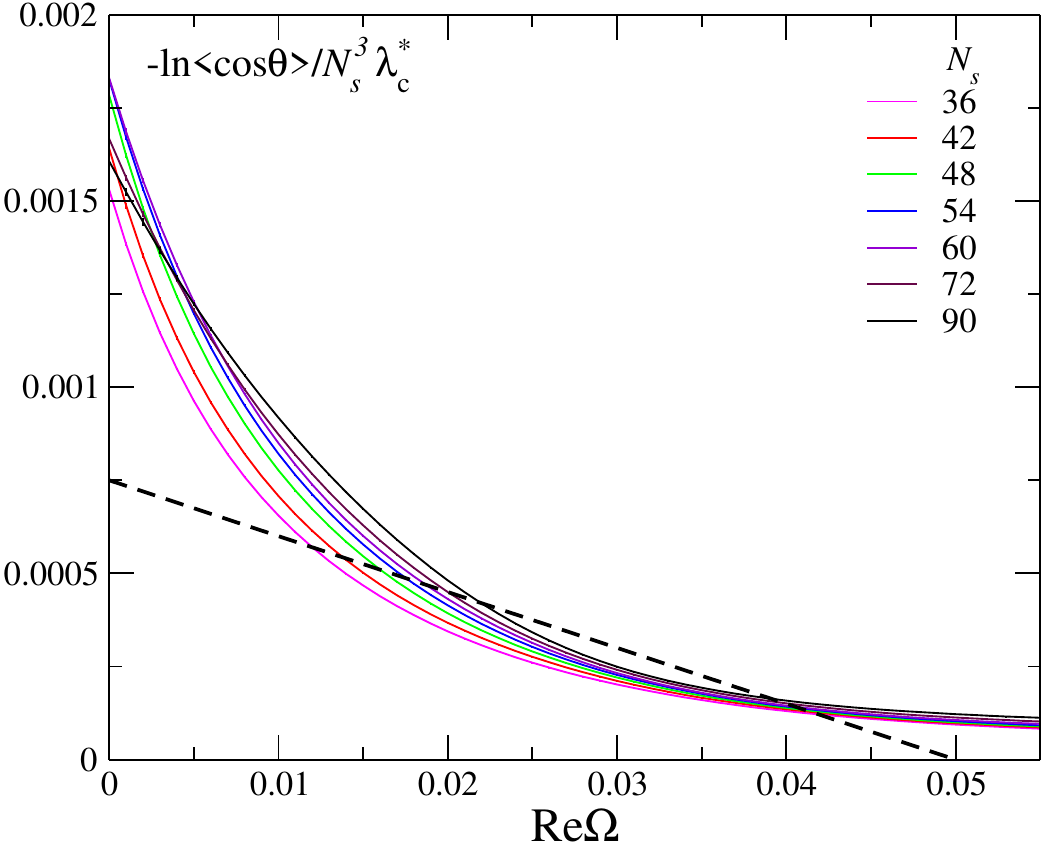}
\vspace{-1mm}
\caption{Left: $\langle \theta^n \rangle_c/ (n! N_s^3 \lambda^*_c)$ and $\ln \langle \cos \theta \rangle/ (N_s^3 \lambda^*_c)$ classified by ${\rm Re} \hat{\Omega}$ on a $90^3 \times 6$ lattice.
Right: Spatial volume dependence of $\ln \langle \cos \theta \rangle/ (N_s^3 \lambda^*_c) $ for $N_s=36$ --90.}
\label{fig3}
\end{figure}

Using the critical value $\lambda^*_c$ of Eq.~(\ref{eq:keff3f}), the phase at the critical point is
$\theta \approx N_s^3 \lambda^*_c \tanh \left( \mu/T \right) {\rm Im} \hat{\Omega}$.
Since $\tanh \left( \mu/T \right) < 1$, 
when $\mu/T$ is increased along the critical line, the phase fluctuation does not increase.
We compute $\langle \cos \theta \rangle_{{\rm Re} \hat{\Omega} =x}$ for each value of ${\rm Re} \hat{\Omega}$ on the configurations used in Ref.~\cite{Kitazawa2023pos}. 
At the same time, we calculate 
$\langle \theta^n \rangle_c = [N_s^3 \lambda^*_c \tanh \left( \mu/T \right)]^n \langle ({\rm Im} \hat{\Omega})^n \rangle_c$, and investigate the convergence properties of the cumulant expansion. 
The lattice size is $N_t=6$, $N_s=36,$ 42, 48, 54, 60, 72, and 90.
Figure~\ref{fig3} (left) shows $\langle \theta^2 \rangle_c$, $\langle \theta^4 \rangle_c$, $\langle \theta^6 \rangle_c$, and $\ln \langle \cos \theta \rangle$ as functions of ${\rm Re} \hat{\Omega}$ for the maximum phase fluctuation case $\tanh \left( \mu/T \right) =1$ with $N_s =90$.
When we classify by the values of ${\rm Re} \hat{\Omega}$ and take the average, we find that the expectation values are almost independent of the simulation parameters $\beta^*$ and $\lambda^*$.
Therefore, we take the average over $\beta^*$ and $\lambda^*$ of the configurations.
This figure indicates that the second-order cumulant is dominant, i.e., 
$\ln \langle \cos \theta \rangle \approx \langle \theta^2 \rangle_c/2$.
Although this cumulant expansion generally may not converge for large $\theta$, this result is due to $\lambda^*_c$ being sufficiently small.
Since the variance $\langle \theta^2 \rangle_c$ is dominant, $\ln \langle \cos \theta \rangle$ increases in proportion to the volume $N_s^3$.
In Fig.~\ref{fig3}, we plot $\ln \langle \cos \theta \rangle / (N_s^3 \lambda^*_c)$. 
The volume dependence is shown in Fig.~\ref{fig3} (right) for $N_s=36$ -- 90. 
They do not change much depending on the volume as expected.

To roughly evaluate the effect of complex phase around the critical point, 
we approximate $\langle \cos \theta \rangle$ with the dashed line in Fig.~\ref{fig3} (right): 
$\ln \langle \cos \theta \rangle / (N_s^3 \lambda^*_c) \approx d_0 + d_1 {\rm Re} \hat{\Omega}$ with $d_0=0.00075$ and $d_1=-0.015$ for the case of $\tanh \left( \mu/T \right) =1$.
Then, 
we get
\begin{eqnarray}
\langle \cos \theta \rangle_{{\rm Re} \hat{\Omega}} |\det M|^{N_{\rm f}}
\approx \exp \left[ \ln \langle \cos \theta \rangle_{{\rm Re} \hat{\Omega}} + N_s^3 \lambda^*_c {\rm Re} \hat{\Omega} \right]
\approx e^{N_s^3 \lambda^*_c d_0} \exp \left[ N_s^3 \lambda^*_c (1-d_1) {\rm Re} \hat{\Omega} \right]
\label{eq:phaseeffl}
\end{eqnarray}
Since $e^{N_s^3 \lambda^*_c d_0}$ does not affect the calculation of expectation values, multiplying $w(x)$ by $\langle \cos \theta \rangle$ in Eq.~(\ref{eq:zpl}) is the same as replacing $\lambda^*_c$ of the phase quenched QCD with $\lambda^*_c(1-d_1)$.
Even when $\tanh \left( \mu/T \right) =1$, $d_1$ is about $-0.015$. 
If the effect of complex phase is added, $\lambda^*_c \sim \kappa_c^{N_t}$ will be reduced by at most $1.5\%$.
Therefore, for $N_t=6$, the complex phase reduces $\kappa_c$ by $0.25\%$.
The change from the phase quenched QCD is about the thickness of the lines in the left panel of Fig.~\ref{fig2}.

\section{Conclusions}
\label{summary}

We discussed how to efficiently determine the critical point of 2+1-flavor QCD in the heavy quark region, including the case of finite density.
When the quark determinant is expanded with the hopping parameters $\kappa$, a strong correlation between the expansion terms is observed.
This leads us to a useful approximation for Polyakov-loop-type terms: 
$2L^+(N_t, n) \approx L^0(N_t,n) c_n \hat{\Omega}$. 

If we ignore the complex phase of the quark determinant, using this approximation, we obtain an effective theory with the action of 
$S_{\rm eff} = -6 N_{\rm site} \beta^* \hat{P} - N_s^3 \lambda^* {\rm Re} \hat{\Omega}$.
We performed numerical simulations with this effective theory, and 
we determined the critical value of $\lambda^*$ at which the phase transition changes from first-order to crossover.
Since there are only two parameters, $\beta^*$ and $\lambda^*$, 
once $\lambda^*_c$ is determined at $\mu=0$, 
the critical line at finite $\mu$ in phase quenched QCD can be determined for any $N_{\rm f}$ including the case of $2+1$ flavor QCD
from the relation between $\lambda^*$, $\kappa$, and $\mu$.

Then, by estimating the magnitude of the complex phase of the quark determinant on the critical line at finite $\mu$, we investigated how much the critical line shifts from that of phase quenched QCD.
We found that the shift of the critical line due to the complex phase is very small for $N_t=6$.
The first-order phase transition region becomes narrower exponentially as the chemical potential increases.
We are planning to calculate the critical line at smaller lattice spacings to determine the critical quark mass in the continuum limit.

\paragraph{Acknowledgments} 
This work was supported by JSPS KAKENHI Grant Numbers JP22K03593, JP22K03619, JP21K03550, JP20H01903, JP19H05598, 
This research used computational resources provided by the HPCI System Research project (Project ID: hp220020, hp220024), and SQUID at Cybermedia Center, Osaka University.


\begin{thebibliography}{99}

\bibitem{Saito:2011fs}
H.~Saito, S.~Ejiri, S.~Aoki, T.~Hatsuda, K.~Kanaya, Y.~Maezawa, H.~Ohno, T.~Umeda
[WHOT-QCD],
``Phase structure of finite temperature QCD in the heavy quark region,''
Phys. Rev. D \textbf{84}, 054502 (2011).

\bibitem{Saito:2013vja}
H.~Saito, S.~Ejiri, S.~Aoki, K.~Kanaya, Y.~Nakagawa, H.~Ohno, K.~Okuno and T.~Umeda,
``Histograms in heavy-quark QCD at finite temperature and density,''
Phys. Rev. D \textbf{89}, 034507 (2014).

\bibitem{Ejiri:2019csa}
S.~Ejiri, S.~Itagaki, R.~Iwami, K.~Kanaya, M.~Kitazawa, A.~Kiyohara, M.~Shirogane, T.~Umeda
[WHOT-QCD],
``End point of the first-order phase transition of QCD in the heavy quark region by reweighting from quenched QCD,''
Phys. Rev. D \textbf{101}, 054505 (2020).

\bibitem{Kiyohara:2021smr}
A.~Kiyohara, M.~Kitazawa, S.~Ejiri and K.~Kanaya,
``Finite-size scaling around the critical point in the heavy quark region of QCD,''
Phys. Rev. D \textbf{104}, 114509 (2021).

\bibitem{Cuteri:2020yke}
F.~Cuteri, O.~Philipsen, A.~Sch\"on and A.~Sciarra,
``Deconfinement critical point of lattice QCD with $N_f$=2 Wilson fermions,''
Phys. Rev. D \textbf{103}, 014513 (2021).

\bibitem{Wakabayashi:2021eye}
N.~Wakabayashi, S.~Ejiri, K.~Kanaya and M.~Kitazawa,
``Scope and convergence of the hopping parameter expansion in finite-temperature quantum chromodynamics with heavy quarks around the critical point,''
PTEP \textbf{2022}, 033B05 (2022)

\bibitem{Kitazawa2023pos}
M.~Kitazawa, R.~Ashikawa, S.~Ejiri, K.~Kanaya, H.~Sugawara, 
``Critical point in heavy-quark region of QCD on fine lattices,'' 
PoS \textbf{LATTICE2023}, 190 (2024).

\bibitem{Ejiri:2009hq}
S.~Ejiri, Y.~Maezawa, N.~Ukita, S.~Aoki, T.~Hatsuda, N.~Ishii, K.~Kanaya, T.~Umeda
[WHOT-QCD],
``Equation of State and Heavy-Quark Free Energy at Finite Temperature and Density in Two Flavor Lattice QCD with Wilson Quark Action,''
Phys. Rev. D \textbf{82}, 014508 (2010).

\end{thebibliography}
\end{document}